\documentclass{article}

\usepackage{arxiv}

\usepackage[utf8]{inputenc} 
\usepackage[T1]{fontenc}    
\usepackage{hyperref}       
\usepackage{url}            
\usepackage{booktabs}       
\usepackage{amsfonts}       
\usepackage{nicefrac}       
\usepackage{microtype}      
\usepackage{lipsum}		
\usepackage{graphicx}
\usepackage{natbib}
\usepackage{doi}

\title{A Multi-faceted Semi-Synthetic Dataset for Automated Cyberbullying Detection}

\author{Naveed Ejaz \\
School of Computing, Queen's University, \\
Kingston, ON, Canada \\
\texttt{naveed.ejaz@queensu.ca} \\
\and
Fakhra Razi \\
Department of Computing and Technology\\
Iqra University, Islamabad Campus, Pakistan\\
\texttt{fakhrarazi29914@iqraisb.edu.pk} \\
\and
Salimur Choudhury \\
School of Computing, Queen's University, \\
Kingston, ON, Canada \\
\texttt{salimur.choudhury@queensu.ca}
}

\renewcommand{\shorttitle}{\textit{arXiv} Template}

\hypersetup{
pdftitle={A Multi-faceted Semi-Synthetic Dataset for Automated Cyberbullying Detection},
pdfsubject={q-bio.NC, q-bio.QM},
pdfauthor={David S.~Hippocampus, Elias D.~Striatum},
pdfkeywords={First keyword, Second keyword, More},
}

\begin{document}
\maketitle

\begin{abstract}
In recent years, the rising use of social media has propelled automated cyberbullying detection into a prominent research domain. However, challenges persist due to the absence of a standardized definition and universally accepted datasets. Many researchers now view cyberbullying as a facet of cyberaggression, encompassing factors like repetition, peer relationships, and harmful intent in addition to online aggression. Acquiring comprehensive data reflective of all cyberbullying components from social media networks proves to be a complex task. This paper provides a description of an extensive semi-synthetic cyberbullying dataset that incorporates all of the essential aspects of cyberbullying, including aggression, repetition, peer relationships, and intent to harm. 
The method of creating the dataset is succinctly outlined, and a detailed overview of the publicly accessible dataset is additionally presented. This accompanying data article provides an in-depth look at the dataset, increasing transparency and enabling replication. It also aids in a deeper understanding of the data, supporting broader research use.
 
\end{abstract}


\section{Introduction}
The widespread use of social media has led to an increasing concern about cyberbullying in the contemporary world. Cyberbullying refers to the use of electronic communication, such as social media platforms, messaging apps, or online forums, to harass, threaten, or intimidate individuals \citep{yi2023session}. This pervasive issue has become a growing concern in the digital age, impacting people of all ages, particularly adolescents and young adults \citep{darawsheh2023impact}. Unlike traditional forms of bullying, cyberbullying transcends physical boundaries, making it challenging to detect and address.

In recent years, many methods and datasets have been presented for automated cyberbullying detection. However, the primary concern in the cyberbullying literature is the absence of a widely acknowledged definition \citep{rosa2019automatic}. While some authors consider any exhibition of insulting or offensive behaviour online as cyberbullying, recent studies distinguish between various sub-tasks of cyber aggression \citep{elsafoury2021timeline, mladenovic2021cyberaggression}. Various perspectives suggest that cyberbullying includes elements like the repeated presence of hostile material, peer associations, and a deliberate intention to inflict harm \citep{rosa2019automatic, smith2012cyberbullying,olweus2000bullying}. However, most datasets lack comprehensive coverage of all cyberbullying aspects \citep{ejaz2024comprehensive}. Current datasets frequently categorize a solitary aggressive message exchanged between two users as an instance of cyberbullying. Within these datasets, the instructions for annotators commonly centre on recognizing cyberbullying by emphasizing aggressive or offensive content, often neglecting the inclusion of other factors that contribute to cyberbullying.  Therefore, current datasets limit in-depth research on cyberbullying, especially in real-life environments. 

In \citep{ejaz2024comprehensive}, we presented a detailed methodology for creating a semi-synthetic dataset. This study introduced a novel framework to develop a comprehensive dataset addressing the existing research gap in cyberbullying. The dataset was meticulously crafted by amalgamating all four dimensions of cyberbullying. Text messages were extracted from an authentic dataset, complemented by the generation of synthetic user data. This synthesized information ensured the inclusion of diverse message exchanges occurring randomly among distinct user pairs, thereby incorporating the crucial element of repetition. Furthermore, the framework included a novel metric, termed "peerness," designed to quantify the likelihood of two users being considered peers. To gauge the intent of harm, a numeric value was systematically calculated by considering the ratios of aggression and repetition. Consequently, the proposed dataset encompassed all facets of cyberbullying, presenting a unique compilation of repeated aggressive messages exchanged between users, accompanied by quantitative measures of both the degree of peerness and the intent to cause harm.

This accompanying data article complements the primary research \citep{ejaz2024comprehensive} by offering comprehensive insights into the dataset. It enhances transparency, facilitates replicability, and promotes thorough data comprehension. Moreover, the currently presented dataset is an extended version compared to the initial prototype dataset presented in \citep{ejaz2024comprehensive}. Table \ref{Spec_table} outlines key details related to the dataset.

\begin{table}[h]
  \centering
  \caption{Data Specifications Table}
  \label{Spec_table}
  \begin{tabular}{l  p{9cm}}
    \hline
    \textbf{Subject} & Artificial Intelligence \\
    \textbf{Specific Subject Area} & Social Media Analysis, Human-computer interaction, Natural Language Processing \\
    \textbf{Data Format} & Raw, Analyzed \\
    \textbf{Type of Data} & Tables \\
    \textbf{Data Collection} & Initially, a dataset is created by gathering text messages from diverse sources comprising aggressive and non-aggressive content. Additionally, a synthetic dataset is created using user data. These datasets are utilized to generate user-to-user messages, incorporating repetition. The users' dataset helps assess the level of similarity between each user pair. The degree of harmful intent is determined based on the ratio of aggression and repetition. The detection of cyberbullying occurs when the interaction between users includes substantial levels of aggression, repetition, peerness, and intent to cause harm. \\
    \textbf{Data Accessibility} & Repository name: Mendeley Data \\
    & Data Identification Number: 10.17632/wmx9jj2htd.2 \\
    & Direct URL to Data: \url{https://data.mendeley.com/datasets/wmx9jj2htd/2} \\
    \textbf{Related Research Article} & N. Ejaz, F. Razi, and S. Choudhury, "Towards comprehensive cyberbullying detection: A dataset incorporating aggressive texts, repetition, peerness, and intent to harm," Computers in Human Behavior, vol. 153, pp. 108123, 2024. doi: \href{https://doi.org/10.1016/j.chb.2023.108123}{10.1016/j.chb.2023.108123}. [1] \\
    \hline
  \end{tabular}
\end{table}
The following key aspects highlight the value of the dataset in the realm of cyberbullying.

\begin{itemize}

    \item The proposed dataset covers different aspects of cyberbullying, like aggression, repetition, peer involvement, and the intention to cause harm. Integrating all these factors helps gain a thorough understanding of cyberbullying.
    
    \item The dataset quantifies the subjective metrics of "peerness" and "intent to harm." These measurable indicators provide researchers with a standardized framework and thus pave the path for consistent and comparable evaluations across various studies.

    \item There is a lack of a universal definition of Cyberbullying. While some researchers consider “occurrence among peers” as an essential condition of cyberbullying, others disagree. A similar disagreement exists between the factors like repetition and intent to harm. To cater to these varying definitions, the dataset has been made customizable by providing separate scores for each constituent element of cyberbullying. The researchers can add or remove a certain ingredient of cyberbullying to adapt the dataset for different definitions. 
    
    \item Researchers can utilize the dataset for training and evaluating their cyberbullying detection models. This dataset becomes valuable in comparative studies, allowing researchers to evaluate the strengths and weaknesses of various cyberbullying detection approaches.
    
    \item Researchers can enhance this dataset by incorporating communication exchanges involving individuals from diverse backgrounds. The provision of code for generating synthetic data facilitates straightforward dataset extension. Furthermore, researchers may choose to include additional components of cyberbullying, such as power imbalances.

\end{itemize}

\section{EXPERIMENTAL DESIGN, MATERIALS AND METHODS}
\label{sec:headings}
The proposed data set is semi-synthetic. Initially, The users' data is generated synthetically. Let's define a set $U$  where each element $u_i \in U$ represents a unique user. Each user $u_i$ is associated with a tuple of attributes: $ u_i = (User_i, Age_i, Gender_i, SchoolName_i, Grade_i)$ where $User_i$ is a unique identifier for user $i$, $Age_i$ is the age of user $i$, $Gender_i$ is the gender of user $i$, $SchoolName_i$  is the name of the school attended by user $i$ and $Grade_i$ is the grade of user $i$ in school. This data generation process creates a dataset for 100 users featuring distinctive attributes. Each user $u_i$  is assigned a unique UserID ranging from $1$ to $100$. The age of each individual $Age_i$ is randomly chosen, spanning from 8 to 18 years, with gender randomly assigned as Male, Female, or Others. Additionally, $SchoolName_i$ are randomly picked from a pool of 15 schools, labelled School1 to School15. The $Grade_i$ of a particular user is determined based on $Age_i$  by assuming that a student of age six joins grade 1. 

The peerness value quantifies the peer relationship between two users. For the mutual peerless values, we define a matrix $P = [p_{ij}]$ where  $0 \leq p_{ij} \leq 1$ where each element $p_{ij}$ represents the peerness value between user $i$ and user $j$. $p_{ij}$  is the peerness value between $User_ i$ and $User_j$. The peerness value is symmetric, implying $p_{ij} = p_{ji}$. Based on the users’ data, the pairwise peerness values between each pair of users $U_i$ and $U_j$ are generated by the method described in \cite{ejaz2024comprehensive}. A value of peerness close to 1 denotes higher peerness, whereas a value near 0 signifies less shared attributes or characteristics between the users.

Aggressive and non-aggressive messages were compiled from diverse sources \cite{elsafoury2020cyberbullying,kumar2018aggression}.  $A = \{a_1, a_2, a_3, \ldots, a_n\} $ is the set of all collected aggressive messages where each  $a_i \in A $ represents an individual aggressive message, and $n$ is the total number of aggressive messages. $N = \{n_1, n_2, n_3, \ldots, n_m\}$  is the set of all non-aggressive messages, where each $n_j \in N $ represents an individual non-aggressive message and $m$  is the total number of non-aggressive messages. Using the sets $U$, $A$, and $N$, the communication data $C$ was generated randomly. Each element $c_i \in C$ is represented by the tuple $c_i = (Date_i, Time_i, User_i, User_j, Message_{i,j}, Label_{i,j})$, where $Date_i$ is the date when the message was sent, $Time_i$ is the time when the message was sent, $User_i$  is the identifier of the sender, $User_j$ is the identifier of the receiver, $Message_{i,j}$ is the content of the message sent, $Label_{i,j}$ is a binary label with $1$ indicating the message is aggressive and $0$ indicating the message is not aggressive. Random dates and times were assigned to each message.

The frequency of messages was calculated for each pair of users using the synthetically generated conversational data. This frequency of messages is termed repetition. The quantified value of intent to harm is computed based on normalized values of total aggressive messages and repetition. Let $D$ be the set of all interaction instances between users, where each element $d_i \in D $ is represented by the tuple: $d_i = (User_i, User_j, TotalMessages_{i,j}, AggressiveCount_{i,j}, IntentToHarm_{i,j}, Peerness_{i,j}, CB\_Label_{i,j})$, $User_i$ is the identifier of the sender, $User_j$ is the identifier of the receiver, $TotalMessages_{i,j}$ is the total number of messages sent from $User_i$ to $User_j$, $AggressiveCount_{i,j}$ is the number of aggressive messages sent by User1 to User2. $IntentToHarm_{i,j}$ is a value between $0$ and $1$ indicating the intent to harm in the interaction \cite{ejaz2024comprehensive}. $Peerness_{i,j}$ is a value between $0$ and $1$ indicating the peerness level between $User_i$ and $User_j$.  $CB\_Label_{i,j}$ is a binary label, with $1$ indicating cyberbullying and $0$ indicating non-cyberbullying.

\begin{figure}
  \centering
  \includegraphics[width=\linewidth]{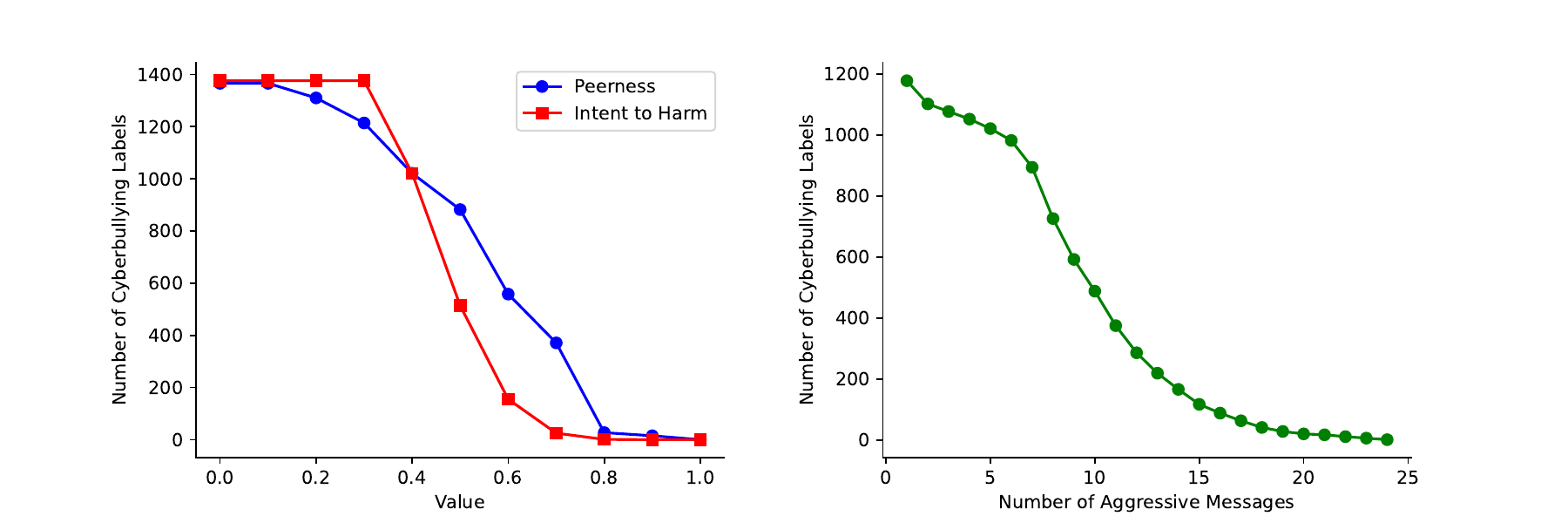}
  \caption{Peerness, Intent to Harm and Number of Aggressive Messages vs Number of Cyberbullying Labels}
  \label{fig:figure1}
\end{figure}

Cyberbullying label $CB\_Label_{i,j}$ between $User_i$ and $User_j$ are determined by applying thresholds to the respective values of $Peerness_{i,j}$, $TotalMessages_{i,j}$, and $IntentToHarm_{i,j}$. Figure \ref{fig:figure1} delineates the alterations in these thresholds vis-à-vis the quantity of selected cyberbullying messages. To illustrate the variations in the threshold of a specific parameter, the values of the other two thresholds are maintained at their initially designated magnitudes.  The presented dataset was created by applying a 0.4 threshold for both peerness and intent to harm, along with the inclusion of five repeated messages.


\section{Data Description}
There are six CSV files in the dataset. The first file, “1. users\_data.csv”,  encompasses synthetically generated data from 100 students, including their user IDs, ages, genders, school names and grades. The second file, “2. peerness\_values.csv,” shows the peerness values between users. There are a total of 4950 values corresponding to each pair of users. The peerness values are between 0 and 1, where a value close to 1 indicates more likeliness of two users to be peers and vice versa. The third and fourth files, denoted as "3. Aggressive\_All.csv" and "4. Non\_Aggressive\_All.csv," amalgamate all aggressive and non-aggressive messages. These files are provided separately so that the researchers can train and test the aggression detection models separately and then employ them in the context of cyberbullying. The fifth file with the name “5. Communication\_Data\_Among\_Users.csv” encompasses communication exchange messages between different users, detailing the date, time, user IDs, message content, and an aggression label (1= Aggression, 0=non-aggression). This is the second important file in the dataset, which lists the communication among different users. The last file, named “6. CB\_Labels.csv,” is the most important file, which lists the total and aggressive messages a user sends to a particular user. For each pair of users engaged in communication, this file provides corresponding peerness values, intent-to-arm scores, and cyberbullying labels. 

Table 1 presents the key characteristics of the dataset. It comprises 100 users, with 118,828 aggressive messages and an equivalent number of non-aggressive messages. The dataset encompasses 90,356 communication messages, spanning 9,512 unique pairs of users engaged in conversations. Among these, there are 30,777 aggressive messages exchanged between all users. The dataset includes 1,021 conversations identified as instances of cyberbullying, while the remaining 8,490 conversations are categorized as non-cyberbullying interactions. It is to be noted that this dataset is an updated version of the prototype dataset development discussed in \cite{ejaz2024comprehensive}. As posited in \cite{ejaz2024comprehensive}, the thresholds about peerness and intent to harm were set to 0.4, while the minimum requisite count for aggressive messages was determined to be 5. 

\begin{table}[h]
  \centering
  \caption{Dataset Characteristics}
  \begin{tabular}{lr}
    \toprule
    \textbf{Characteristic} & \textbf{Value} \\
    \midrule
    Total Users & 100 \\
    Total Aggressive Messages in Collection & 118,828 \\
    Total Non-Aggressive Messages in Collection & 118,828 \\
    Total Communication Messages & 90,356 \\
    Total Conversations between unique pairs of users & 9,512 \\
    Total Aggressive Messages between all users & 30,777 \\
    No. of Cyberbullying Conversations & 1,021 \\
    No. of Non-Cyberbullying Conversations & 8,490 \\
    \bottomrule
  \end{tabular}
\end{table}

Table 2, excerpted from “CB\_Labels.csv”, provides a sample of 6 rows from the final dataset. The table comprises columns such as "User 1" and "User 2," denoting the identification numbers of the involved users. "No. Of Messages" signifies the total messages exchanged in the corresponding conversation, while "No. of Aggressive Messages" delineates the count of messages displaying aggressive content. The columns "Intent to Harm" and "Peerness" provide numerical representations, ranging from 0 to 1, indicating the perceived intent to cause harm and the level of peerness between User 1 and User 2 in communication, respectively. The "Cyberbullying Label" column assigns a binary label (0 or 1), designating whether the conversation is categorized as cyberbullying (1) or not (0). 

\begin{table}[h]
  \centering
  \caption{An excerpt from CB\_Labels.csv}
  \begin{tabular}{ccccccc}
    \toprule
    \textbf{User 1} & \textbf{User 2} & \textbf{No. of Messages} & \textbf{No. of Aggressive Messages} & \textbf{Intent to Harm} & \textbf{Peeriness} & \textbf{Cyberbullying Label} \\
    \midrule
    1 & 20 & 10 & 2 & 0.22 & 0.57 & 0 \\
    2 & 49 & 33 & 18 & 0.68 & 0.77 & 1 \\
    12 & 52 & 4 & 1 & 0.175 & 0.7 & 0 \\
    14 & 32 & 24 & 11 & 0.53 & 0.63 & 1 \\
    29 & 77 & 15 & 6 & 0.39 & 0.43 & 0 \\
    30 & 34 & 23 & 7 & 0.44 & 0.63 & 1 \\
    \bottomrule
  \end{tabular}
\end{table}

\section{Conclusions and Future Work}

This data article associated with \cite{ejaz2024comprehensive} describes the details of a publicly available semi-synthetic dataset addressing the cyberbullying research gap. The process of generating the dataset is briefly explained, along with a detailed description of the details of the dataset. The described dataset comprehensively covers all facets of cyberbullying, presenting a distinctive compilation of repeated aggressive messages exchanged between users, complemented by quantitative measures of both peerness and intent to cause harm. This accompanying data article complements the primary research, providing in-depth insights into the dataset creation process. It contributes to transparency, supports replicability, and aids in a thorough understanding of the data. 

To ensure the continuity and relevance of conversations, future datasets will aim for a more contextual message selection process that preserves the natural flow of dialogue. This will help create more realistic interactions and improve the model's ability to detect nuances in conversational patterns. Efforts will be made to diversify the user data beyond the initial student-only focus, enabling the application of our findings to a wider array of demographics. This expansion will contribute to a more comprehensive understanding of cyberbullying across various age groups and social contexts. The subjectivity and context-dependency inherent in determining threshold values for elements like Peerness, Repetition, and Intent to Harm will be critically examined. Future research will explore systematic methods for setting these thresholds, thereby reducing potential biases and improving the model's robustness. Investigating a range of threshold settings will also be prioritized to ensure that the selection of these parameters is optimized for both accuracy and relevance in different scenarios. Recognizing the complexity of cyberbullying behaviour, subsequent work will focus on developing a multi-class classification system to replace the current binary approach. This will enable a more granular detection of cyberbullying types, enriching our understanding of the phenomenon and improving the support and intervention strategies

\bibliographystyle{unsrtnat}
\bibliography{references}  






\end{document}